\documentclass[onecolumn,preprintnumbers,amsmath,amssymb]{revtex4}
\usepackage{graphicx}
\usepackage{latexsym}
\usepackage{dcolumn}
\usepackage{stmaryrd}

\usepackage{graphicx}
\usepackage{dcolumn}
\usepackage{bm}

\usepackage{bm}

\begin{document}
\newcommand{\eq}{\begin{equation}}                                                                         
\newcommand{\eqe}{\end{equation}}             

\title{Self-similar solutions of the G-equation - analytic description 
of the flame surface}

\author{ I. F. Barna}
\address{KFKI Atomic Energy Research Institute of the Hungarian Academy 
of Sciences, \\ (KFKI-AEKI), H-1525 Budapest, P.O. Box 49, Hungary}

\date{\today}


\begin{abstract} 
The main feature of the flame kinematics can be desribed with the G-equation. 
We investigate the solutions of the G-equation with the well-known self-similar Ansatz. The results are discussed and the method how to get self-similar solutions is briefly mentioned. 
  
\end{abstract}

\maketitle
To understand the dynamics of flame fronts is a general interest for combusion
engineering. One way to gain such knowledge is the investigation of the G-equation model. 
Since the G-equation was introduced by Markstein in 1964 \cite{mark} numerous analytical \cite{bondar} and numerical approaches
\cite{pitsch,long,dekena} have been used to find the solutions. 
The study of \cite{bondar} uses the method of characteristics for nonlinear 
first order partial-differential equations(PDA). 

Dekena \cite{dekena} uses a computaional fluid dynamics(CFD) code Fire to investigate turbulent flame propagation in Gasoline Direct Injection engines. Further references for application of the G-equation is given therein. 

In the G-equation model the flame itself is treated as a surface (flame front) that separates the burnt from the unburnt gas. A detailed derivation of the model can be found in \cite{bondar}. The flame front is described here by the relation of
\eq
G(r,z,t) = z - \zeta(r,t) = 0. 
\eqe
The most general form of the G-equation is the following: 
\eq
\frac{\partial \zeta(r,t)}{\partial t} + u \frac{\partial \zeta(r,t)}{\partial r}  -v +
 S_l\sqrt{1+  \left(  \frac{\partial \zeta(r,t)}{\partial r} \right)^2}  = 0.   
\label{eque}
\eqe  
Where $u(r,z,t)$ and $v(r,z,t)$ are the radial and the axial components of the gas velocity, 
and $S_L$ is the constant modulus of the laminar burning velocity,  
respectively.
                                                             
In the following we use a well-known method, which is very popular to 
investigate nonlinear PDAs, namely searching for self-similar solutions. 
We are looking for solution of (\ref{eque}) in the form of 
\eq
\zeta(r,t)=t^{-\alpha}f\left(\frac{r}{t^\beta}\right):=t^{-\alpha}f(\eta).
\label{ans}
\eqe
The similarity exponents $\alpha$ and $\beta$ are of primary physical importance since 
$\alpha$  represents the rate of decay of the magnitude $\zeta(x,t)$, 
while $\beta$  is the rate of spread 
(or contraction if  $\beta<0$ ) of the space distribution as time goes on. 
This kind of solutions have a broad range of applicability, e.g. a large number of diffusion-reaction-type equations \cite{robi} or time-dependent telegraph-type heat-conduction equation \cite{imi} can be 
examined as well. 

Substituting this into (\ref{eque}) we have
\begin{eqnarray}
-\alpha t^{-\alpha -1}f(\eta) - \beta t^{-\alpha-\beta-1}f'(\eta)r - 
u(r,z,t)t^{-\alpha-\beta}f'(\eta) - v(r,z,t) +  \\ \nonumber 
 S_l\sqrt{1+ t^{-2\alpha-2\beta}[f'(\eta)]^2 }  = 0 
\label{3}
\end{eqnarray}
where prime denotes differentiation with respect to $\eta.$ 

After setting  set $u=S_l=1$ and $v=0$,
one can see that this is a non-linear ordinary differential equation(ODE)
if and only if $\alpha  = -1$ and $\beta = 1$ ({\it {the universality relation}}).
The corresponding ODE we shall deal with is
\eq
f(\eta) - \eta f'(\eta) - f'(\eta) + \sqrt{1+ [f'(\eta)^2]} = 0.  
\label{diff}
\eqe
The solutions can be easily obtained and read: 
\eq
f(\eta)_1 = c_1\eta - c_1 - \sqrt{1+c_1^2}; \hspace{1.1cm} f(\eta)_2 = c_1\sqrt{\eta(\eta-2)}
\eqe
where $c_1$ is the real integration constant. 
The solutions can be seen in Fig 1. 
The corresponding self-similar solutions are  
\eq
\zeta(r,t)_1 =  c_1r - c_1 - \sqrt{1+c_1^2};  \hspace{1.1cm}     \zeta(r,t)_2 = c_1 t \sqrt{ \left( \frac{r(r-2t)}{t^2}  \right) }.
\eqe
Both solutions can be prooven with direct derivation.  The first solution is trivial 
however the non-trivial second one is presented on Fig. 2. 

We may consider more sophysticated flow systems where thew radaial and axial componenst of the 
gas velocity and the $S_L$ are functions of time and radial position. 
First let's consider that the velocities and the constant modulus of the laminar burning velocity  are 
real numbers ($u,v,S_l \> \epsilon \> \Re$.)
Now the ODE has the following form 
\eq
f(\eta) - \eta f'(\eta) - uf'(\eta) - v + S_l\sqrt{1+ [f'(\eta)^2]} = 0.   
\label{diff}
\eqe
The solutions became a bit more complicated 
\eq
f(\eta)_1 =  c_1\eta + u c_1 + v - S_l\sqrt{1+c_1^2};  \hspace{1.1cm}     
f(\eta)_2 = c_1\sqrt{-u^2-\eta^2 +S_L^2 - 2\eta u} + v. 
\eqe 
The function $f(\eta)_2$ is presented in Fig. 3. with the following fixed set of 
parameters $(u = 2, v=-1, c_1=1, S_L =3 ). $ 
Note, that the solution has now a compact support with a non-vanishing first derivatives 
at the border. This means that there is no flux-conservation at the boundaries. 
There are more analytical solutions for special $u(x,t), v(x,t)$ and $S_l(x,t)$ functions. The only remaining task is to give reasonable 
physical interpretation for $u(x,t), v(x,t)$ and $S_l(x,t)$.

Another generaly interesting question is the dispersion relation and the attenuation distances. It can be examined how wave equations or other nonlinear evolutionary equations propagate plain waves in time and in space.  
Inserting the usual plain wave approximation $\zeta(r,t) = e^{i(kr + \omega t)}$ into
(\ref{eque}) the dispersion relation and the attenuation distance can be obtained.
These are the followings
\eq
v_p =  \frac{\omega}{Re(k)} = \frac{\omega}{ \frac{1}{2} \sqrt{|-1+k^2|} 
(1- signum[-1+k^2])};
\eqe
\eq
\alpha = \frac{1}{Im(k)} = \frac{1}{\omega + k +\frac{1}{2}\sqrt{|-1+k^2|}  (1- signum[-1+k^2]) } 
\eqe
inserting the relation of $\omega = k c$ where $c$ is the propagation velocity of the signal the formulas only depend on the wavenumber vector k. 
Considering $c=1$ propagation speed Fig. 4. shows the phase velocity as the function of the wavenumber. 
The $(1- signum[-1+k^2]) $ in the formula is responsible for the compact support of the function.  
Fig. 5. shows the attenuation distance of the various waves. 
At $k=1$ the $\alpha(k)$ function is non-analytic. \\ 
In this short study we just wanted to present that self-similar solution can be easily used to generate analytic solutions for the G-equation. With more general, and more physical relations for the 
radial and axial gas flow velocity hopefully more physical solutions can be obtained. Anyhow, any kind of analytical solution of a nonlinear 
PDA can be usefull giving a solid basis for testing complex and 
sophisticated two or three dimensional 
numerical finite-element computational fluid dynamics codes. 


\newpage


\begin{figure}
\scalebox{0.4}{
\rotatebox{0}{\includegraphics{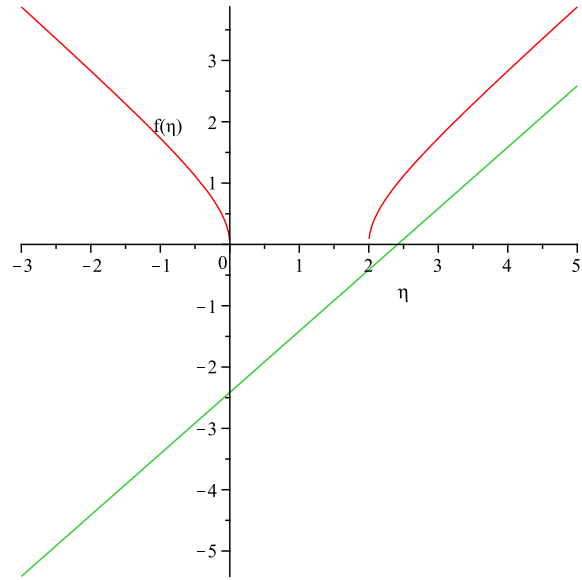}}}
\vspace*{0.4cm}
\caption{Solutions of Eq. (5),  
red line presents $f(\eta)_2$ and the green one shows 
$f(\eta)_1$
 .}
\label{fig:r}       
\end{figure} 

\begin{figure}
\scalebox{0.4}{
\rotatebox{0}{\includegraphics{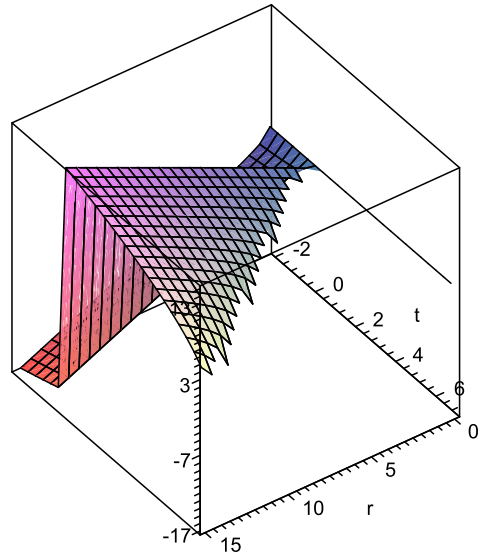}}}
\vspace*{0.4cm}
\caption{The self-similar solution  
$\zeta(r,t)_2$, from Eq. (7) in the range r = 0 .. 15, t = -2 .. 7 for
$c_1 = 1 $.} 
\label{fig:r}       
\end{figure} 

\begin{figure}
\scalebox{0.4}{
\rotatebox{0}{\includegraphics{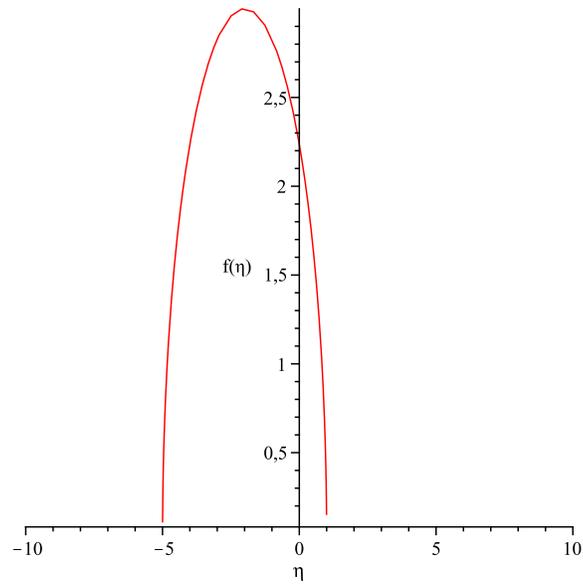}}}
\vspace*{0.4cm}
\caption{The solution $f(\eta)_2, \>\>   u = 2, v=-1, c_1=1, S_L =3 .$} 
\label{fig:r}       
\end{figure}

\begin{figure}
\scalebox{0.4}{
\rotatebox{0}{\includegraphics{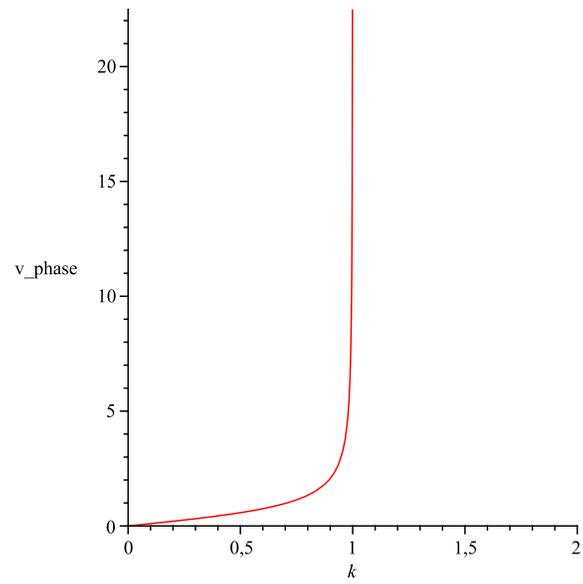}}}
\vspace*{0.4cm}
\caption{The dispersion relation $v_p(k) $ for Eq. (2).} 
\label{fig:r}       
\end{figure} 

\begin{figure}
\scalebox{0.4}{
\rotatebox{0}{\includegraphics{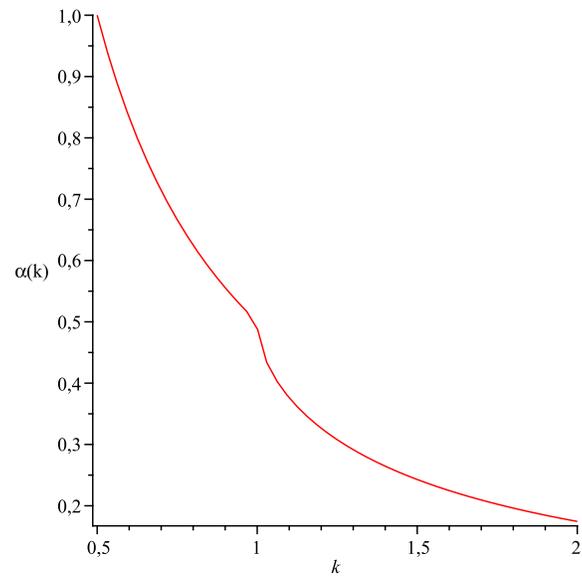}}}
\vspace*{0.4cm}
\caption{The attenuation function $\alpha(k)$ for Eq. (2).} 
\label{fig:r}       
\end{figure} 

\end{document}